# Assessment of mobility deficit and treatment efficacy in adhesive capsulitis by measurement of kinematic parameters using IMU sensors

Miloš Ajčević *
*Department of Engineering and Architecture, University of Trieste*
Trieste, Italy
majcevic@units.it

Manuela Deodato *
*Department of Medicine Surgery and Health Sciences; Department of Life Sciences, University of Trieste*
Trieste, Italy
mdeodato@units.it

Luigi Murena
*Department of Medicine Surgery and Health Sciences, University of Trieste*
Trieste, Italy
lmurena@units.it

Aleksandar Miladinović
*Department of Engineering and Architecture, University of Trieste*
Trieste, Italy
aleksandar.miladinovic@phd.units.it

Susanna Mezzarobba
*Department of Medicine Surgery and Health Sciences, University of Trieste*
Trieste, Italy
mezzarob@units.it

Agostino Accardo
*Department of Engineering and Architecture, University of Trieste*
Trieste, Italy
accardo@units.it

*Abstract*—There is a growing research interest towards the use of wireless IMU sensors to assess disability, monitor progress and provide feedback to patients on range of motion and movement performance during upper body rehabilitation. The quality of movement in patients with adhesive capsulitis and relative treatment efficacy has not yet been studied using inertial and magnetic sensors. The aim of this study was to investigate the possibility to quantitatively evaluate capsulate-related deficit versus healthy controls and to assess treatment efficacy by measurement of shoulder kinematic parameters with ISEO protocol using inertial and magnetic measurement system technology. We enrolled 6 patients with adhesive capsulitis (AC) who underwent the experimental assessment by using a set of wireless IMU sensors at the baseline (T0) and after the 15 one-hour individual sessions of physiotherapy (T1). The range of motion in elevation, abduction and the scapulo-humeral rhythm kinematic parameters were extracted from measurements performed in enrolled AC patients and in 7 healthy controls. The results of this preliminary study showed that proposed approach based on measurement of shoulder kinematic parameters with ISEO protocol using IMU wireless sensors can be useful in mobility deficit assessment of patients with adhesive capsulitis, as well as for monitoring of treatment efficacy and its further personalization.

I. INTRODUCTION

The shoulder is a complex of joints that allow the relative motion of the humerus respect to the thorax. Traumatic events or degenerative pathologies can cause an improper balance of shoulder structures leading to musculoskeletal diseases. Nowadays, shoulder pathologies are an increasing problem among workers and overhead athletes, as well as in the general population [1], [2]. Shoulder pain is the third most common musculoskeletal dysfunction and together with upper limb dysfunctions reduces self-care and functional autonomy. Rehabilitation treatments generally aim to recover arm function and reduce shoulder pain, to increase the quality of life of patients [3].

There is a growing research interest towards the development and evaluation of wearable systems, as well as for protocols for their use in upper body rehabilitation [4]-[7]. Indeed, currently accelerometers and inertial measurement units - IMUs are most commonly used to assess disability, monitor progress and provide feedback to patients on range of motion and movement performance during upper body rehabilitation [4].

Primary adhesive capsulitis is reported with a prevalence from 2% to 5.3% of the general population [8], [9]. Patients show a progressive painful restriction of the passive and active range of motion in all directions of the gleno-humeral joint. In this condition the capsular subsynovial layer could lead to chronic inflammation, thickening and fibrosis. For this reason, adhesive capsulitis is commonly called *frozen shoulder* [10], [11]. The conservative therapy of the adhesive capsulitis includes physical therapy, intra-articular injections, and oral anti-inflammatory medications. When these therapies fail, surgical intervention is indicated [8], [10], [12].

Appropriate clinical management strategies are based on the effectiveness of specific treatments. For this, clinicians need accurate and reliable methods to evaluate pain and functional limitations. Usually, the most common measurement tools for clinical evaluation of shoulder performance are mechanical manual systems such as the goniometer, inclinometer and plurimeter. However, mechanical measurement tools have some limitations and provide little information about the quality of movements or the ability to carry out functional tasks [13], [14].

Beside the on range of motion a clinical parameter heavily affected in most shoulder disorders is the scapulohumeral rhythm (SHR) [15], which is the coordinated and coupled movement between scapula and humerus, when the latter is elevated [16]. From the clinical viewpoint, the SHR is mainly analyzed during humerus elevations in the sagittal, scapular and frontal plane [15]. Optimal scapular function plays a pivotal roll in alignment and function of the shoulder joints. The scapulo-humeral rhythm allows the proper motion for functional shoulder tasks. Scapular dyskinesis has been observed in different shoulder syndromes such as impingement syndrome, shoulder instability and adhesive capsulitis. For this reason, the assessment of the scapula motion and the result of treatments should be considered as a basic aspect in the objective examination of shoulder performance [17]. Other valid systems used for human motion analysis include vision-based systems and optoelectronic systems. However, they are relatively expensive and restricted to laboratory settings, and require specific technical expertise [13], [14]. On the other hand, the inertial sensors offer practical alternatives relatively cheaper and portable for kinematic and functional assessment. Inertial sensors incorporate accelerometers and gyroscopes to assess movement control, coordination and fluency through acceleration and angular velocity.

*These authors contributed equally
The work is partially supported by Master's in Clinical Engineering - University of Trieste.

It is possible to differentiate pathological and healthy shoulders movement [14]. Recently a motion analysis protocol named ISEO (INAIL Shoulder & Elbow Outpatient-clinic protocol) was proposed to assess the upper-limb kinematic, included scapulo-humeral rhythm, with inertial and magnetic sensors [6].

The quality of movement in patients with adhesive capsulitis and relative treatment efficacy has not yet been studied using inertial and magnetic sensors.

The aim of this study is to investigate the possibility to quantitatively evaluate capsulate-related deficit versus healthy controls and to assess treatment efficacy by measurement shoulder kinematic parameters with ISEO protocol using inertial and magnetic measurement system technology in patients with adhesive capsulitis.

## II. MATERIALS AND METHODS

### A. Study population and protocol

A total of 13 subjects were enrolled, 6 patients (3F/3M; age 53.8±4.3 years) with adhesive capsulitis (AC) diagnosis and 7 healthy controls (HC) (3F/4M; age 41.3±4.3).

Patients with AC aged between 18 and 65 were enrolled after diagnosis by an experienced orthopedic physician. The exclusion criteria were: pregnancy; serious psychiatric pathologies; significant surgical procedures during the previous 12 months; contraindications of corticosteroid infiltrative therapy or rehabilitation treatment; serious pathologies such as traumas, tumors, or infections; physical therapy or other conservative treatments in the previous 3 months; contraindications relating to physiotherapy. The study was carried out over a period of 5 months from May to October 2019 at Physiotherapy Unit of Trieste University Hospital. The research was conducted in accordance with the Declaration of Helsinki, all patients signed the informed consent and the privacy rights of all subjects were be observed.

The enrolled patients underwent the experimental assessment by using a set of inertial sensors at the baseline (T0) and after the 15 one-hour individual sessions of physiotherapy (T1). In particular, the range of motion in elevation, abduction and the scapulo-humeral rhythm were assessed by measurements performed using inertial sensors according to the ISEO protocol [6] and by subsequent signal and data analysis.

In addition, the range of motion in elevation, abduction and the scapulo-humeral rhythm were also measured applying the same measurement and analysis method in 7 healthy subjects as a HC group.

### B. ISEO protocol and signal acquisition

The kinematic signals and parameters were acquired by using MTw wireless sensor units (Xsens Technologies, NL). Each MTw sensor unit contains a 3D-gyroscope, accelerometer and magnetometer, which together provide the orientation of the technical coordinate system of the MTw relative to a global, earthbased coordinate system. ISEO protocol was applied according to the following procedure [6]. Sensor were placed on thorax, scapula and humerus (Fig. 1). The thorax MTw was placed on the flat portion of the sternum. The scapular MTw was placed on the skin, just above the scapular spine, over the central third between the angulus acromialis and the trigonum spinae, aligning the MTw with the upper edge of the scapular spine. The upper arm MTw was placed on an elastic cuff, over the central third of the humerus, slightly posterior.

A static measure is performed with the subject standing in a pre-defined posture to complete the sensor-to-segment calibration, i.e. the calculation of the anatomical coordinate systems: upright position, elbow flexed at 90°.

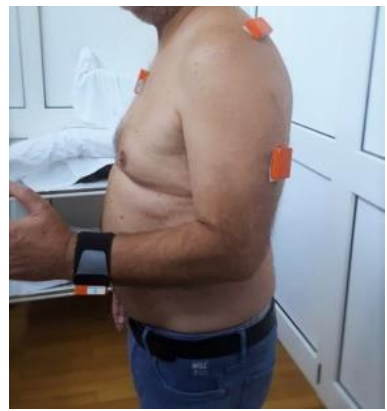

**Fig.1** Wireless IMU sensor units placement according to ISEO protocol.

### C. Treatment protocol

The Physical Therapy Protocol was organized in 15 one-hour individual sessions over a period of three months. An integrated treatment of manual therapy and active exercises was performed. As regards to manual therapy, in consisted of: micro-mobilization of accessory joints: clavicula (Fig.2a) acromioclavicular and sternoclavicular joints, scapula (Fig.2b) and humeral head (Fig.2c); cervical (Fig.2d) and dorsal rachis (Fig.2e) that are involved in the biomechanics of shoulder; finally combined multidirectional mobilization technique (Fig.2f) was performed in the gleno-humeral joint [18]. Concerning active exercise counter-resistance mobilizations with post-isometric release in anterior flexion, abduction, external rotation and internal rotation (Fig.2g) and postural active exercises (Fig.2i) were performed.

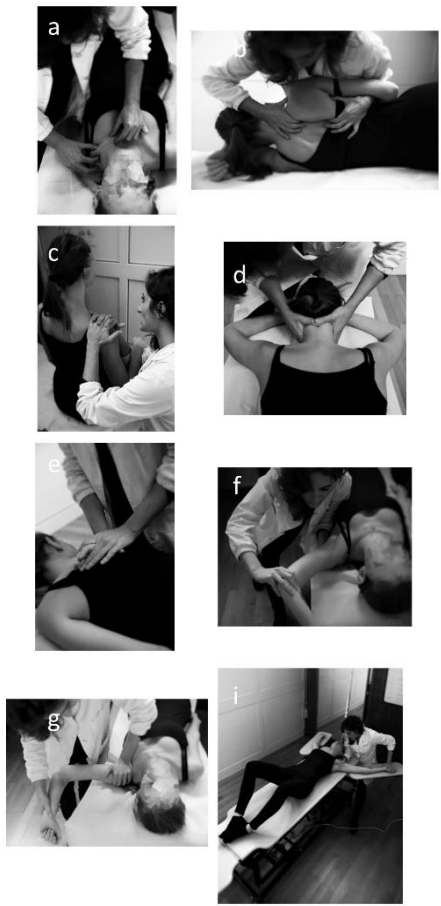

**Fig.2** Physical Therapy Protocol. (a) Clavicula mobilization; (b) Scapula mobilization; (c) Micro-mobilization of humeral head; (d) Cervical mobilization; (e) Dorsal mobilization; (f) Combined multidirectional mobilization; (g) Counter-resistance mobilizations; (i) Postural active exercise.

*D. Kinematic parameters analysis*

The elevation range of motion (ROMe) was obtained from the tracings using the sensors placed on the wrist and thorax as a reference and expressed in degrees (°).

Regarding the scapulohumeral-rhythm, measured during a shoulder abduction task performed in the frontal plane (Fig3.), the humeral range of motion in abduction (ROMa), the range of motion of scapula (ROMs) were extracted.

ROMa was measured using the sensors placed on the humerus and thorax reference, while ROMs was identified by sensors placed on spine of the scapula and the thorax. In addition, the activation time of the scapula and humerus during the same movement were extracted.

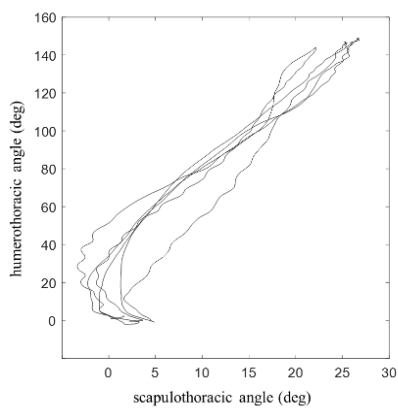

**Fig.3** Scapulohumeral rhythm measured during a shoulder abduction task performed in the frontal plane. The repetitions of the movement performed by the subject during the task are reported.

The differences between the AC and HC were investigated by using Mann-Whitney U-test, while the differences pre- and post-treatment were investigated Wilcoxon signed rank test.

## III. RESULTS

The extracted values of range of motion in elevation (ROMe) and abduction (ROMa) observed in AC patients at the baseline assessment and after the treatment are reported in Table 1. Table 2 reports the ROMe and ROMa measured in healthy controls. As expected, ROMe and ROMa were lower in AC patients at baseline than those assessed in HC, although the difference was not statistically significant (p=0.56 and p=0.10, respectively). After the treatment, there was a slight improvement of ROMe in 4 out of 6 patients. The improvement of ROMa was more enhanced and observed in 5 out of 6 patients, in which ROMa reached HC levels. In average, elevation increased from 145.6±8.7° to 148.2±13.8°, abduction increased from 125.7 ±30.3° to 163.4 ±26.5°, but both not statistically significant (p=0.28 and p=0.15, respectively). Mean±SD values of ROM of the scapula observed in HC was 26.1±7.8°. In Table 3 are reported the values of ROMs in AC patients at baseline and after the treatment. ROMs improved significantly after the treatment from 21.0±7.1° to 34.6±7.7°, (p=0.03). Mean±SD values of activation times were 1.81±0.93 and 1.93±1.09 for scapula and humerus in HC, respectively. Activation times of the scapula and humerus during abduction are reported in Table 4. After the treatment scapulo-humeral rhythm improved the coordinated activation time of humerus from 3.60±1.11s to 3.03±1.71s.

TABLE I. RANGE OF MOTION IN ELEVATION AND ABDUCTION OBSERVED IN AC AT BASELINE ASSESSMENT AND POST-TRATMENT

| Patient | ROMe (°) | | ROMa (°) | |
|---|---|---|---|---|
| | T0 | T1 | T0 | T1 |
| 1 | 135.8 | 151.3 | 83.6 | 178.3 |
| 2 | 160.3 | 170.0 | 133.7 | 199.9 |
| 3 | 148.3 | 151.3 | 148.9 | 153.3 |
| 4 | 145.4 | 127.8 | 155.0 | 160.2 |
| 5 | 145.3 | 143.0 | 140.6 | 120.6 |
| 6 | 138.3 | 146.1 | 92.2 | 168.0 |
| Mean±SD | 145.6±7.9 | 148.3±12.5 | 125.7±27.6 | 163.4±24.2 |

TABLE II. RANGE OF MOTION IN ELEVATION AND ABDUCTION OBSERVED IN HC

| Subject | ROMe (°) | ROMa (°) |
|---|---|---|
| 1 | 169.2 | 158.4 |
| 2 | 155.6 | 143.1 |
| 3 | 146.9 | 158.1 |
| 4 | 153.1 | 153.6 |
| 5 | 152.6 | 151.4 |
| 6 | 162.8 | 154.4 |
| 7 | 162.8 | 169.7 |
| Mean±SD | 157.6±7.6 | 153.2±5.6 |

TABLE III. RANGE OF MOTION OF SCAPULA IN ABDUCTION OBSERVED IN AC AT BASELINE ASSESSMENT AND POST-TRATMENT

| Patient | ROMs (°) | |
|---|---|---|
| | T0 | T1 |
| 1 | 10.9 | 32.3 |
| 2 | 21.3 | 30.8 |
| 3 | 23.0 | 31.1 |
| 4 | 30.3 | 49.9 |
| 5 | 14.7 | 28.8 |
| 6 | 25.7 | 34.4 |
| Mean ±SD | 21±7.1 | 34.6±7.7 |

TABLE IV. ACTIVATION TIMES OF THE SCAPULA AND HUMERUS DURING ABDUCTION

| Patient | Activ. time scapula (s) | | Activ. time humerus (s) | |
|---|---|---|---|---|
| | T0 | T1 | T0 | T1 |
| 1 | 3.74 | 5.31 | 5.23 | 5.86 |
| 2 | 1.77 | 1.85 | 2.40 | 1.78 |
| 3 | 0.71 | 3.37 | 2.54 | 2.69 |
| 4 | 4.15 | 3.01 | 3.21 | 2.65 |
| 5 | 3.84 | 4.65 | 3.73 | 4.07 |
| 6 | 2.96 | 1.31 | 4.46 | 1.10 |
| Mean ±SD | 2.86±1.36 | 3.25±1.55 | 3.60±1.11 | 3.03±1.71 |

## IV. DISCUSSION

Personalized technology solutions, including advanced wearable devices, should be developed to support the treatment self-empowerment of chronic patients [19]. Patient-specific performance assessment of arm movements is fundamental for rehabilitation therapy [5]. Inertial and magnetic measurement systems are a new generation of motion analysis systems which may diffuse the measurement of upper-limb kinematics ambulatory settings [6].

This study preliminarily assessed the possibility to quantitatively evaluate capsulate-related deficit and relative treatment efficacy, in patients with adhesive capsulitis, by measurement shoulder kinematic parameters with ISEO protocol using inertial and magnetic measurement wireless technology.

Indeed, IMU sensors applied according to the ISEO protocol allowed to assess non-invasively the reduction of mobility in terms of range of motion and scapulohumeral rhythm-related kinematic parameters, between AC patents and HC groups, although statistical differences were not found probably due to limited study sample. In addition, it was possible to monitor quantitatively the treatment progress, to evaluate its efficacy. ROM increased both in elevation (ROMe) and in abduction (ROMa). The greater increase was observed in abduction, in line with recent findings [20]. This suggesting that an integrated type of physical therapy including manual therapy and active exercise may be useful in patients with AC in term of ROM. However, at the end of physiotherapy protocol statistical differences were not found in shoulder elevation and in shoulder abduction between AC and HC groups. These results should be interpreted taking into account that fifteen one-hour individual sessions of physiotherapy are few to obtain some more significant improvements in AC patients.

Finally, we assessed the alteration scapulohumeral rhythm in adhesive capsulitis patients. In patients with shoulder pathologies, there is abnormal activation of the scapula in association with the movements of the glenohumeral joint, both in terms of the amount of movement and activation timing [17]. The data obtained from the evaluation of the patients in our sample are in line with these results, demonstrating an early activation of the scapula compared to the humerus in patients with adhesive capsulitis. After the treatment a better synchronization in terms of activation times of scapula and humerus was observed, as well as a significant increase of the ROM of scapula (ROMs). We suppose it can be linked to the increase in the pain threshold of the levator muscle of the scapula, but also to an improvement in the scapulohumeral rhythm at the end of the treatment. In fact, it was observed that at the baseline all AC patients had an alteration of this rhythm in terms of higher activation times and early activation of the scapula compared to the humerus, while at the end of treatment the movement pattern in terms of coordination of the scapula and humerus activations was found to be similar to the healthy controls. This alteration of the rhythm, observed in various shoulder pathologies, represents an important factor of pathology onset. It also indicates that a rehabilitation treatment not only at the level of the glenohumeral joint, but also involving the neighboring districts can be more effective in restoring the correct bachelor-humeral rhythm.

## V. CONCLUSION

In conclusion, the results of this preliminary study showed that the proposed approach based on measurement of shoulder kinematic parameters with ISEO protocol using IMU wireless sensors can be useful in mobility deficit assessment of patients with adhesive capsulitis, as well as for monitoring of treatment efficacy and its further personalization.

ACKNOWLEDGMENT

The authors would like to thank all the patients and volunteers, as well as the staff of Physiotherapy Unit of Trieste University Hospital for the participation in the study.